\documentstyle[multicol,aps,pre,twocolumn]{revtex}
\input{psfig}
\newcommand{\bm}[1]{\mbox{\boldmath $#1$}}
\begin{document}
\twocolumn[\hsize\textwidth\columnwidth\hsize\csname@twocolumnfalse\endcsname

\title{Lagrangian method for multiple correlations in passive scalar advection}
\author{U.~Frisch$^{1}$, A.~Mazzino$^{1,2}$,
A.~Noullez$^{1}$ and M.~Vergassola$^{1}$\\
\small{$^{1}$ CNRS, Observatoire de la C\^ote d'Azur, B.P. 4229,
06304 Nice Cedex 4, France.}\\
\small$^2$ INFM--Dipartimento di Fisica, Universit\`a di Genova, I--16146
Genova, Italy.}
\draft
\date{\today}
\maketitle
\begin{abstract}
A Lagrangian method is introduced for calculating simultaneous
$n$-point correlations of a passive scalar advected by a random
velocity field, with random forcing and finite molecular diffusivity
$\kappa$. The method, which is here presented in detail, is
particularly well suited for studying the $\kappa\to 0$ limit when the
velocity field is not smooth. Efficient Monte Carlo simulations based
on this method are applied to the Kraichnan model of passive scalar
and lead to accurate determinations of the anomalous intermittency
corrections in the fourth-order structure function as a function of
the scaling exponent $\xi$ of the velocity field in two and three
dimensions. Anomalous corrections are found to vanish in the limits $\xi \to 0$
and $\xi \to 2$, as predicted by perturbation theory.
\end{abstract}
\pacs{PACS number(s)\,: 47.10.+g, 47.27.-i, 05.40.+j}]

\section{Introduction}
\label{sec:intro}

Robert Kraichnan's model of passive scalar advection by a
white-in-time velocity field has been particularly fertile ground for
theoreticians trying to develop a theory of intermittency
\cite{K94,GK95,CFKL95} (see also Refs.~\cite{G98,CK98}). Although the
model leads to closed equations for multiple-point moments, only
second-order moments can be obtained in closed analytic form
\cite{K68}. Theoretical predictions differed as to the behavior of
higher-order quantities, regarding in particular the survival or the
vanishing of intermittency corrections (anomalies) in certain
limits.  Obtaining reliable numerical results was thus an important
challenge.  Until recently numerical simulations have been based on
the direct integration of the passive scalar partial differential
equation and have been limited to two dimensions\,\cite{CK98,KYC95,FGLP97}.
Such calculations are  delicate; to wit, the difficulty of
observing for the second-order structure function the known 
high-P\'eclet number asymptotic
scaling \cite{K68}. Also, the numerical scheme used in
Refs.~\cite{CK98,KYC95} involves a slightly
anisotropic velocity field which is not expected to give exactly the
right scaling laws for the passive scalar \cite{FW96}.

Lagrangian methods for tackling the Kraichnan model and which require
only the integration of ordinary differential equations were recently
proposed independently by Frisch, Mazzino and Vergassola \cite{FMV98} and by
Gat, Procaccia and Zeitak \cite{GPZ98}. Our goal here is to give a detailed
presentation of the Lagrangian method and to present new results.

In Section~\ref{sec:lag} we give the theoretical background of the
Lagrangian method for general random velocity fields which need not be
white-in-time. In Section~\ref{sec:kappazero} we investigate the limit
of vanishing molecular diffusivity which depends crucially on how
nearby Lagrangian trajectories separate. We then turn to the Kraichnan
model which is given a Lagrangian formulation
(Section~\ref{sec:kraichnan}). Then, we show how it can be solved
numerically by a Monte-Carlo method (Section~\ref{sec:num}) and
present results in both two and three space dimensions (Section~\ref{sec:res}).
We make concluding remarks in Section~\ref{sec:conclusion}.

\section{The Lagrangian method}
\label{sec:lag}

The (Eulerian) dynamics of a passive scalar field $\theta({\bm r},t)$
advected by a velocity field ${\bm v}({\bm r},t)$ is described by
the following partial differential equation (written in space
dimension $d$)\,:
\begin{equation}
\label{passive}
\partial_t\theta({\bm r},t)+{\bm v}({\bm r},t)\cdot\nabla\,
\theta({\bm r},t)=\kappa\nabla^2\theta({\bm r},t)+f({\bm r},t) ,
\end{equation}
where $f({\bm r},t)$ is an  external source (forcing) of scalar
and $\kappa$ is the molecular diffusivity. In all that follows we
shall assume that $\theta({\bm r},t)=0$ at some distant time in the past
$t=-T$ (eventually, we shall let $T\to \infty$).

In this section the advecting velocity ${\bm v}$ and the forcing $f$
can be either deterministic or random. In the latter case, no
particular assumption is made regarding their statistical properties.

In order to illustrate the basic idea of the Lagrangian strategy, let
us first set $\kappa=0$. We may then
integrate (\ref{passive}) along its characteristics, the Lagrangian
trajectories of tracer particles, to obtain
\begin{equation}
\theta({\bm r},t)=\int^t_{-T} f({\bm a}(s;{\bm r},t),s)\,ds,
\label{formalli}
\end{equation}
where ${\bm a}(s;{\bm r},t)$ is the position at time $s\le t$ of the fluid
particle which will be at position ${\bm r}$ at time
$t$. (Two-time Lagrangian positions of this type were also
used in Kraichnan's Lagrangian History Direct Interaction theory
\cite{K65}.) This Lagrangian position, which will henceforth be denoted
just ${\bm a}(s)$, satisfies the ordinary differential equation
\begin{equation}
{d{\bm a}(s)\over ds}={\bm v}\left({\bm a}(s),s\right),
\qquad {\bm a}(t)={\bm r}.
\label{lagnok}
\end{equation}

For $\kappa>0$ we shall now give a stochastic generalization of this
Lagrangian representation. Roughly, $\theta$ will be the average of a
random field $\phi$ which satisfies an advection--forcing equation
with no diffusion term, in which the advecting velocity is the sum of
${\bm v}$ and a suitable white-noise velocity generating Brownian diffusion.

To be more specific we need to introduce some notation. We shall use a
set of $n$ $d$-dimensional time-dependent random vectors
\begin{equation}
\dot{\bm w}_i(s)= \{\dot w_{i,\alpha};\,\,i=1,\ldots,n;\,\,
\alpha=1,\ldots,d\},
\label{defwi}
\end{equation}
which are Gaussian, identically distributed, independent of each other and
independent of both ${\bm v}$ and $f$. The time dependence is assumed
to be white noise\,:
\begin{equation}
\left\langle\dot w_{i,\alpha}(s)\dot w_{j,\beta}(s')
\right\rangle_w= \delta_{ij}\delta_{\alpha\beta}\delta(s-s').
\label{wn}
\end{equation}
The notation $\langle \cdot\rangle_w$ stands for ``average over the
$\dot{\bm  w}_i$'s for a fixed realization of  ${\bm v}$ and $f$''.
Similarly, $\langle \cdot\rangle_{vf}$ stands for ``average over ${\bm
v}$ and $f$ for a fixed realization of the $\dot{\bm  w}_i$'s''.
Unconditional averages are denoted just $\langle\cdot\rangle$. Clearly,
\begin{equation}
\langle\cdot\rangle=\left\langle \langle
\cdot\rangle_w\right\rangle_{vf}
= \left\langle \langle \cdot\rangle_{vf}\right\rangle_{w}.
\label{averages}
\end{equation}
The white noise, which is a random distribution, is here denoted by
$\dot w(s)$ since it is the time derivative of the Brownian motion (or
Wiener--L\'evy) process. In the numerical implementation we shall work
with increments of $w(s)$.

We can now state the main result which is at the basis of our
Lagrangian method.

\vspace{1mm}
{\em Let $\phi_i({\bm r},t)$ ($i=1,\ldots,n$) be the solutions
of the following advection--forcing equations\,:}
\begin{eqnarray}
&&\partial_t \phi_i({\bm r},t)+\left({\bm v}({\bm r},t)+\sqrt{2\kappa}
\dot{\bm w}_i(t)\right)\cdot\nabla \phi_i({\bm r},t)= f({\bm
r},t).\nonumber\\
&&\phi_i({\bm r},-T)=0, \qquad
i=1,2,\ldots,n.\label{advect-forc}
\end{eqnarray}
{\em For any ${\bm r}_1$, ${\bm r}_2$, ${\bm r}_3$, \ldots, we have}
\begin{eqnarray}
&&\theta({\bm r_1})=\left\langle \phi({\bm r_1})\right\rangle_w,
\quad \forall i\label{waverag1}\\
&&\theta({\bm r_1})\theta({\bm r_2})=\left\langle
\phi_i({\bm r_1})\phi_j({\bm r_2})\right\rangle_w,\,\,\,\forall
i\neq j, \label{waverag2}\\
&&\theta({\bm r_1})\theta({\bm r_2})\theta({\bm r_3})=
\left\langle \phi_i({\bm r_1})\phi_j({\bm r_2})\phi_k({\bm
r_3})\right\rangle_w, \nonumber\\
&&\forall i,j,k,\,\,\,\, i\neq j\neq k \neq i,\label{waverag3}\\
&&..................................................,\nonumber
\end{eqnarray}
{\em where all the fields $\theta$ and $\phi$ are evaluated at the same
time $t$.}

The proof of (\ref{waverag1}) is obtained by
taking the mean value of (\ref{advect-forc}) (only over $w$) and
noting that
\begin{equation}
\left\langle \sqrt{2\kappa}\dot {\bm w}_i\cdot\nabla
\phi_i\right\rangle_w= -\kappa \nabla^2 \left\langle \phi_i\right\rangle_w.
\label{novikov}
\end{equation}
(This is a standard result for linear stochastic equations having
white-noise coefficients; it is derived using Gaussian integration by
parts (see Ref.~\cite{F95}, Chap.~4). For similar derivations see
Refs.~\cite{K68,BF74,FW97}.)

To prove (\ref{waverag2}), we first derive from
(\ref{advect-forc}), assuming $i\neq j$,
\begin{eqnarray}
&&\partial_t\left(\phi_i({\bm r}_1)\phi_j({\bm r}_2)\right) + [{\bm
v}({\bm r}_1)\cdot\nabla_1+{\bm v}({\bm r}_2)\cdot\nabla_2\nonumber\\
&& +\sqrt{2\kappa}(\dot{\bm w}_i\cdot\nabla_1+\dot{\bm
w}_j\cdot\nabla_2)]\phi_i({\bm r}_1)\phi_j({\bm r}_2) = \nonumber\\
&&f({\bm r}_1)\phi_j({\bm r}_2) + f({\bm r}_2)\phi_j({\bm r}_1),
\label{dtphiphi}
\end{eqnarray}
where $\nabla_1$ and $\nabla_2$ stand for $\nabla_{{\bm r}_1}$
and $\nabla_{{\bm r}_2}$. We then average (\ref{dtphiphi}) (over $w$)
and use a relation similar to (\ref{novikov})
\begin{eqnarray}
&&\left\langle \sqrt{2\kappa}\left(\dot {\bm w}_i\cdot\nabla_1
+ \dot {\bm w}_2\cdot\nabla_2\right)\phi_i({\bm r}_1)\phi_j({\bm
r}_2)\right\rangle_w\nonumber\\
&&= -\kappa \left(\nabla^2_1 + \nabla^2_2\right)
\left\langle \phi_i({\bm r}_1)\phi_j({\bm r}_2)\right\rangle_w.
\label{novikov2}
\end{eqnarray}
(Notice that cross terms involving $\nabla_1\cdot\nabla_2$ disappear
because of the independence of $\dot{\bm w}_i$ and $\dot{\bm w}_j$.)
The higher order equations in (\ref{waverag3}) and following are
proved similarly.

Because of the absence of a diffusion operator in (\ref{advect-forc}),
its solution has  an obvious Lagrangian representation
\begin{equation}
\phi_i({\bm r},t)=\int_{-T}^t f({\bm a}_i(s),s)\,ds,
\label{lagrep}
\end{equation}
where
\begin{eqnarray}
&&{d{\bm a}_i(s)\over ds}={\bm v}\left({\bm a}_i(s),s\right)
+\sqrt{2\kappa}\dot{\bm w}_i(s), \label{lag}\\
&&\qquad {\bm a}_i(t)={\bm r}.
\label{lag1}
\end{eqnarray}

So far we have not used the random or deterministic character of ${\bm
v}$ and $f$. In the random case, taking the average of
(\ref{waverag1})--(\ref{waverag3}) over ${\bm v}$ and $f$, we obtain
\begin{eqnarray}
&&\left\langle\theta({\bm r_1})\right\rangle=\left\langle \phi({\bm r_1})\right
\rangle,
\quad \forall i\label{averag1}\\
&&\left\langle\theta({\bm r_1})\theta({\bm r_2})\right\rangle=\left\langle
\phi_i({\bm r_1})\phi_j({\bm r_2})\right\rangle, \,\,\,\forall i\neq j
\label{averag2}\\
&&\left\langle\theta({\bm r_1})\theta({\bm r_2})\theta({\bm r_3})\right\rangle=
\left\langle \phi_i({\bm r_1})\phi_j({\bm r_2})\phi_k({\bm
r_3})\right\rangle, \nonumber\\
&&\forall i,j,k,\,\,\,\, i\neq j\neq k \neq i,\label{averag3}\\
&&.................................................\nonumber
\end{eqnarray}

Eqs.~(\ref{lagrep}), (\ref{lag}), (\ref{lag1}), together with 
(\ref{averag1}), (\ref{averag2}) and higher orders, constitute our
Lagrangian representation for multiple-point moments of the passive
scalar.

Let us stress that, for moments beyond the first order, it is
essential to use more than one white noise. Indeed, we could have made
use of just (\ref{waverag1}) and written
\begin{equation}
\theta({\bm r}_1)\theta({\bm r}_2) =
\langle\phi_i({\bm r}_1)\rangle_w\langle\phi_j({\bm r}_2)\rangle_w,
\quad\forall i,j.
\label{product}
\end{equation}
(Including the case $i=j$.) It is however not possible to
$({\bm v},f)$-average
(\ref{product}) because it involves a {\em product\/} of $w$-averages
rather than just one average as in
(\ref{waverag2}).

In the more restricted context of the Kraichnan model, a functional
integral representation of $n$th order moments involving, as here, $n$
white noises has already been given in Refs.~\cite{BGK97,MC97}.

\section{The limit of vanishing molecular diffusivity}
\label{sec:kappazero}

Interesting pathologies occur when we bring two or more Eulerian space
arguments of the moments to coincidence and {\em simultaneously\/}
let $\kappa\to 0$.

This is already seen on the second-order moment $\langle \theta
^2({\bm r}_1)\rangle$. From (\ref{lagrep}) and (\ref{averag2}), we have
\begin{equation}
\langle \theta^2({\bm r}_1)\rangle=
\int_{-T}^t \int_{-T}^t \left\langle f({\bm a}_i(s),s)f({\bm
a}_j(s'),s')\right\rangle
\, ds ds',
\label{thetac}
\end{equation}
for any $i\neq j$. The differential equations for ${\bm a}_i(s)$ and ${\bm
a}_j(s)$ involve different white noises. If, in the limit $\kappa\to
0$, we simply ignore the $\sqrt{2\kappa}\dot {\bm w}_i$ terms in
(\ref{lag}), we find that all the ${\bm a}_i(s)$'s satisfy the same
equation (\ref{lagnok}) and the same boundary condition ${\bm
a}_i(t)={\bm r}_1$. It is then tempting to conclude that all the ${\bm
a}_i(s)$'s are identical, namely are just the Lagrangian trajectories
${\bm a}(s)$ of the unperturbed ${\bm v}$-flow. As a consequence, we can then
rewrite (\ref{thetac}) as
\begin{eqnarray}
\langle \theta^2({\bm r}_1)\rangle&=&
\int_{-T}^t \int_{-T}^t \left\langle f({\bm a}(s),s)f({\bm
a}(s'),s')\right\rangle
\, ds ds',\nonumber\\
&=& \left\langle\left(\int_{-T}^t f({\bm a}(s),s)\,ds\right)^2\right\rangle.
\label{thetachum}
\end{eqnarray}

For a large class of random forcings $f$ of zero mean the r.h.s. of
(\ref{thetachum}) will diverge $\propto T$ as $T\to \infty$. This is
for example the case when $f$ is homogeneous, stationary and short-
(or delta-) correlated in time as in the Kraichnan model. The reason
of this divergence is that, although the forcing has zero mean, its
integral (along the Lagrangian trajectory) over times long compared to
the correlation time behaves like Brownian motion (in the
$T$-variable) and, thus, has a variance  $\propto T$. From this
na\"{\i}ve procedure we would thus conclude that, when $T=\infty$,
the scalar variance becomes infinite as $\kappa\to 0$.

This conclusion is actually correct when the ${\bm v}$-flow is smooth
(differentiable in the space variable)\,: this is the so-called
Batchelor limit which has been frequently investigated
\cite{K68,BGK97,B59,K74,SS94,CFKL95b}. The conclusion however becomes 
incorrect when the ${\bm v}$-flow is only H\"older continuous,
i.e. its spatial increments over a small distance $\ell$ vary as a
fractional power of $\ell$ (e.g. $\ell^{1/3}$ in Kolmogorov 1941
turbulence). As pointed out in Ref.~\cite{BGK97} (see also
Ref.~\cite{G98}), when ${\bm v}$ is not smooth the solution to
(\ref{lagnok}) lacks uniqueness, so that two Lagrangian particles
which end up at the same point ${\bm r}_1$ at time $t$ may have
different past histories. This is exemplified with the one-dimensional
model
\begin{equation}
{dx\over ds} = - x^{1/3}, \quad -T\le s\le t, \quad x(t)=\epsilon\ge 0,
\label{1dmodel}
\end{equation}
where $x$ is a deputy for ${\bm a}_i -{\bm a}_j$, the separation
between two nearby Lagrangian particles. For $\epsilon>0$, the
solution of (\ref{1dmodel}) is
\begin{equation}
x(s)=\left[\epsilon^{2/3} +{2\over 3}(t-s)\right]^{3/2}.
\label{sol1d}
\end{equation}
If we now set $\epsilon=0$ in (\ref{sol1d}) or consider times $s$ such
that $|t-s| \gg \epsilon^{2/3}$, we obtain
\begin{equation}
x(s)=\left[{2\over 3}(t-s)\right]^{3/2}.
\label{sol1dnoeps}
\end{equation}
This is indeed {\em a\/} solution of (\ref{1dmodel}) with
$\epsilon=0$, but there is another trivial one, namely $x(s)=0$.
Related to this non-uniqueness is the fact that, when $\epsilon$ is
small the solution given by (\ref{sol1d}) becomes independent of
$\epsilon$, namely
is given by the non-trivial solution (\ref{sol1dnoeps}) for
$\epsilon=0$.

Whenever the flow ${\bm v}$ is just H\"older continuous, the separation of
nearby Lagrangian particles proceeds in a similar way, becoming
rapidly independent of the initial separation. Such a law of
separation is much more explosive than would have been obtained for a
smooth flow with sensitive dependence of the Lagrangian
trajectories. In the latter case we would have $x(s)=
\epsilon \,e ^{\lambda(t-s)}$ ($\lambda >0$), which grows exponentially
with $t-s$ but still tends to zero with $\epsilon$.

We propose to call this explosive growth  a Richardson
walk, after Lewis Fry Richardson who was the first to
experimentally observe this rapid separation in turbulent flow and who
was also much interested in the role of non-differentiability in
turbulence \cite{R93}. It is this explosive separation
which prevents the divergence of $\langle \theta ^2({\bm r}_1)\rangle$
when $\kappa\to 0$ (and more generally of moments with several
coinciding points). Indeed, as the time $s$ moves back from $s=t$, even
an infinitesimal amount of molecular diffusion will slightly
separate, say, by an amount $\epsilon$, the Lagrangian particles ${\bm
a}_i(s)$ and ${\bm a}_j(s)$ which coincide at $s=t$. Then, the
Richardson walk will quickly bring the separations to values
independent of $\epsilon$ and, thus of $\kappa$. Hence, the double
integral in (\ref{thetac}), which  involves points ${\bm a}_1(s)$ and ${\bm
a}_2(s)$ with uncorrelated forces when $|s-s'|$ is sufficiently large,
may converge for $T\to \infty$. (For it to
actually converge more specific assumptions must be made about the
space and time correlations of ${\bm v}$ and $f$, which are satisfied,
e.g., in the Kraichnan model.)

An alternative to introducing a small diffusivity is to work at
$\kappa= 0$, with ``point splitting''. For this one replaces
$\langle \theta ^2({\bm r}_1)\rangle$ by $\langle \theta ({\bm
r}_1)\theta ({\bm r'}_1)\rangle$, where ${\bm r}_1$ and  ${\bm r'}_1$
are separated by a distance $\epsilon$. Eventually, $\epsilon \to 0$.
In practical numerical implementations we found that point splitting
works well for second-order moments but is far less efficient than
using a small diffusivity for higher-order moments.

\section{The Kraichnan model}
\label{sec:kraichnan}

The Kraichnan model \cite{K68,K94} is an instance of the passive
scalar equation (\ref{passive}) in which the velocity and the forcing
are Gaussian white noises in their time dependence. This ensures that
the solution is a Markov process in the time variable and that closed
moment equations, sometimes called ``Hopf equations'', can be written
for single-time multiple-space moments such as
$\left\langle\theta({\bm r_1},t)\ldots\theta({\bm
r_n,t})\right\rangle$. The equation for second-order moments was
published for the first time by Kraichnan \cite{K68} and, for
higher-order moments, by Shraiman and Siggia \cite{SS94}. Note that
Hopf's work \cite{H52} dealt with the characteristic functional of
random flow; it had no white-noise process and no closed equations,
making the use of his name not so appropriate in the context of
white-noise linear stochastic equations. The fact that closed moment
equations exist for such problems has been known for a long time (see,
e.g., Ref.~\cite{BF74} and references therein). We shall not need the
moment equations and shall not write them here (for an elementary
derivation, see Ref.~\cite{FW97}).

The precise formulation of the Kraichnan model as used here is the
following.  The velocity field ${\bm v}=\{v_\alpha,
\alpha=1,\ldots,d\}$ appearing in (\ref{passive}) is incompressible,
isotropic, Gaussian, white-noise in time; it has homogeneous
increments with power-law spatial correlations and a scaling exponent
$\xi$ in the range $0< \xi< 2$\,:
\begin{eqnarray}
&&\left\langle [v_\alpha ({\bm r},t)-v_\alpha ({\bm
0},0)] [v_\beta ({\bm r},t)-v_\beta ({\bm 0},0)]
\right\rangle = \nonumber\\ &&2 \delta (t) D_{\alpha\beta} ({\bm r}),
\label{correlations}
\end{eqnarray}
where,
\begin{eqnarray}
D_{\alpha\beta} ({\bm r})= r^{\xi}\left[\left(\xi+d-1\right)
\delta_{\alpha\beta}-\xi\,{r_{\alpha}
r_{\beta}\over r^2}\right].
\label{d-tensor}
\end{eqnarray}
Note that since no infrared cutoff is assumed on the velocity its
integral scale is infinite; this is not a problem since only velocity
increments matter for the dynamics of the passive scalar. Note also
that when a white-in-time velocity is used in (\ref{lag}), the well
known Ito--Stratonovich ambiguity could appear\, \cite{KP92}. This
ambiguity is however absent in our particular case, on account of
incompressibility.

The random forcing $f$ is independent of ${\bm v}$, of zero mean,
isotropic, Gaussian, white-noise in time and homogeneous. Its
covariance is given by\,:
\begin{equation}
\langle f({\bm r},t)\,f({\bm 0},0)\rangle =
F\left(r/L\right)\,\delta(t),
\label{fcorr}
\end{equation} 
with $F(0)>0$ and $F\left(r/L\right)$ decreasing rapidly for $r\gg L$,
where $L$ is the (forcing) integral scale. 

In principle, to be a correlation, the function $F\left(r/L\right)$
should be of positive type, i.e. have a non-negative $d$-dimensional
Fourier transform. In our numerical work we find it convenient to work
with the step function $\Theta_L(r)$ which is equal to unity for $0\le
r\le L$ and to zero otherwise. Hence, the injection rate of passive
scalar variance is $\varepsilon=F(0)/2=1/2$. The fact that the
function $\Theta$ is not of positive type is no problem.  Indeed, let
its Fourier transform be written $E(k)=E_1(k)-E_2(k)$, where
$E(k)=E_1(k)$ whenever $E(k)\ge 0$. Using the step function amounts to
replacing in (\ref{passive}) the real forcing $f$ by the complex
forcing $f_1+i f_2$ where $f_1$ and $f_2$ are independent Gaussian
random functions, white-noise in time and chosen such that their
energy spectra (in the space variable) are respectively $E_1(k)$ and
$E_2(k)$. Since the passive scalar equation is linear, the solution
may itself be written as $\theta_1+i \theta_2$ where $\theta_1$ and
$\theta_2$ are the (independent) solutions of the passive scalar
equations with respective forcing terms $f_1$ and $f_2$. Using the
universality with respect to the functional form of the forcing
\cite{GK96}, it is then easily shown that the scaling laws for the
passive scalar structure functions are the same as for real forcing.

We shall be interested in the passive scalar structure functions of
even order $2n$ (odd order ones vanish by symmetry), defined as
\begin{equation}
S_{2n}(r;L)\equiv \left \langle \left(\theta({\bm r})-\theta({\bm
0})\right)^{2n}\right\rangle.
\label{defsn}
\end{equation}
From Ref.~\cite{K68} (see also Ref.~\cite{FW97}) we know that, for $L\gg
r\gg \eta \sim \kappa^{1/\xi}$, the second-order structure function is
given by
\begin{eqnarray}
S_2(r;L) = C_2 \,\varepsilon \,r^{\zeta_2},\quad \zeta_2 = 2-\xi,
\label{s2inrange}
\end{eqnarray}
where $C_2$ is a dimensionless numerical constant. If there were no 
anomalies, we would have, for $n>1$,
\begin{equation}
S_{2n}(r;L) = C_{2n}\, \varepsilon^n\, r^{n\zeta_2}.
\label{snnoanom}
\end{equation}
Note that (\ref{s2inrange}) and (\ref{snnoanom}) do not involve the
integral scale $L$.
Actually, for $n>1$, we have {\em anomalous\/} scaling with
$S_{2n}(r;L)\propto r^{\zeta_{2n}}$ and $\zeta_{2n}<n\zeta_2$. More
precisely, we have
\begin{eqnarray}
&&S_{2n}(r;L) = C_{2n}\, \varepsilon^n\, r^{n\zeta_2}\,
\left({L\over r}\right)^{\zeta^{\rm anom}_{2n}},\\
&&\zeta^{\rm anom}_{2n} \equiv n\zeta_2 - \zeta_{2n},
\label{snanom}
\end{eqnarray}
where the structure function now displays a dependence on the integral
scale $L$.  Our strategy will be to measure the dependence of
$S_{2n}(r;L)$ on $L$ while the separation $r$ and the injection rate
$\varepsilon$ are kept fixed and, thereby, to have a direct
measurement of the anomaly $\zeta^{\rm anom}_{2n}$.

Let us show that, in principle, this can be done by the Lagrangian
method of Section~\ref{sec:lag} using $2n$ tracer (Lagrangian)
particles whose trajectories satisfy (\ref{lag}). The structure
function of order $2n$ can be written as a linear combination of
moments of the form $\left\langle \theta({\bm r}_1)\theta({\bm
r}_2)\ldots \theta({\bm r}_{2n})\right\rangle$, where $p$ of the
points are at location ${\bm r}$ and $2n-p$ at location ${\bm 0}$
($p=0,\ldots,2n$). Because of the symmetries of the problem, $p$ and
$2n-p$ give the same contribution. Thus, we need to work only with the
$n+1$ {\em configurations\/} corresponding to  $p=0,\ldots,n$. For
example, we have
\begin{equation}
S_4(r;L)= 2 \left\langle \theta ^4({\bm r})\right\rangle 
-8\left\langle \theta ^3({\bm r})\theta({\bm 0})\right\rangle
+6\left\langle \theta ^2({\bm r})\theta ^2({\bm 0})\right\rangle.
\label{examples4}
\end{equation}
Let us first consider the case of the two-point function (second-order
moment). Using (\ref{lagrep}), (\ref{averag2}) and the independence of
${\bm v}$ and $f$, we have
\begin{eqnarray}
&&\left\langle \theta({\bm r}_1)\theta({\bm
r}_2)\right\rangle=\nonumber\\
&&\left\langle\int_{-T}^t\int_{-T}^t \left\langle f({\bm
a}_1(s_1),s_1)f({\bm a}_2(s_2),s_2)\right\rangle_f\,ds_1ds_2\right\rangle_{vw}.
\label{tt1}
\end{eqnarray}
Here, $\langle \cdot \rangle_f$ is an average over the forcing and
$\langle \cdot \rangle_{vw}$ denotes averaging over the velocity and
the $\dot{\bm w}$'s, and ${\bm a}_1(s_1)$ and ${\bm a}_2(s_2)$ satisfy
(\ref{lag}) with the ``final'' conditions ${\bm a}_1(t) = {\bm r}_1$and
${\bm a}_2(t) = {\bm r}_2$, respectively.

In (\ref{tt1}) the averaging over $f$ can be carried out explicitly
using  (\ref{fcorr}). With our step-function choice for  $F$, we obtain
\begin{equation}
\langle \theta({\bm r}_1)\,\theta({\bm r}_2)\rangle= \langle T_{12}(L)
\rangle_v,
\label{c2}
\end{equation}
where
\begin{equation}
T_{12}(L) = \int_{-T}^{t} \Theta_L(|{\bm a}_1(s) - {\bm a}_2(s)|)\; ds
\label{time}
\end{equation}
is the amount of time that two tracer particles arriving  at ${\bm r}_1$ and
${\bm r}_2$ and moving backwards in time spent with their mutual
distance $|{\bm a}_1(s)-{\bm a}_2(s)| < L$. Whether the particles move
backwards or forward in time is actually irrelevant for the Kraichnan model
since the velocity, being Gaussian, is invariant under reversal.

For the four-point function, we proceed similarly and use the Wick
rules to write fourth-order moments of $f$ in terms of sums of  products of
second-order moments, obtaining 
\begin{eqnarray}
\langle \theta({\bm r}_1)\theta({\bm r}_2)\theta({\bm r}_3)\theta({\bm
r}_4)\rangle &=& \langle T_{12}(L) T_{34}(L) \rangle_{vw} \nonumber \\
&+& \langle T_{13}(L) T_{24}(L) \rangle_{vw} \nonumber \\ &+& \langle
T_{14}(L) T_{23}(L) \rangle_{vw}.
\label{c4}
\end{eqnarray}
Expressions similar to Eqs.~(\ref{c2}) and (\ref{c4}) are easily
derived for higher order  correlations.

We see that the evaluation of structure functions and moments has been
reduced to studying certain statistical properties of the random time
that pairs of particles spend with their mutual distances less than
the integral scale $L$.  Generally, the distance between pairs of
particles tends to increase with the time elapsed but, occasionally,
particles may come very close and stay so; this will be the source of
the anomalies in the scaling.

\section{Numerical implementation of the Lagrangian method}
\label{sec:num}

In Section~\ref{sec:kraichnan} we have shown that structure functions
of the passive scalar are expressible in terms of ${\bm v}$-averages
of products of factors $T_{ij}(L)$. For the structure function of
order $2n$, these products involve configurations of $2n$ particles,
$p$ of which end at locations ${\bm r}$ at time $s=t$ and the
remaining $2n-p$ at location ${\bm 0}$. In the Kraichnan model ${\bm
v}$ and $f$ are stationary, so that after relaxation of transients,
$\theta$ also becomes stationary. We may thus calculate our structure
functions at $t=0$. Time-reversal invariance of the ${\bm v}$ field
and of the Lagrangian equations allows us to run the $s$-time forward
rather than backward. Also, $T_{ij}(L)$ is sensitive only to differences in
particle separations, whose evolution depends only on the difference
of the velocities at ${\bm a}_i$ and ${\bm a}_j$. Furthermore, the
${\bm v}$ field has homogeneous increments. All this allows us to work
with $2n-1$ particle separations, namely
\begin{eqnarray}
&&\tilde{\bm a}_i(s) \equiv {\bm a}_i(s) -{\bm a}_{2n}(s), \quad
i=1,\ldots,2n-1,\label{defatilde}\\
&&\tilde{\bm a}_i(t)= \tilde{\bm r}_i  \equiv {\bm r}_i-{\bm r}_{2n}.
\label{defrtilde}
\end{eqnarray}
Using (\ref{lag}) we find that the quantities $\tilde{\bm a}_i(s)$
satisfy $2n-1$ (vector-valued) differential equations which involve
the  differences of velocities 
${\bm v}({\bm a}_i(s),s)-{\bm v}({\bm a}_{2n}(s),s)$. The statistical
properties of the solutions remain unchanged if we subtract 
${\bm a}_{2n}(s)$ from all the space arguments. We thus obtain the
following Lagrangian equations of motion for the $2n-1$ particle separations\,:
\begin{eqnarray}
&&{d\tilde{\bm a}_i(s)\over ds}=\tilde{\bm v}_i(s)
+\sqrt{2\kappa}\tilde{\bm w}_i(s)
\label{lagtilde}\\
&&\tilde{\bm v}_i(s)  \equiv {\bm v}(\tilde{\bm a}_i(s),s)-{\bm
v}({\bm 0},s),\label{defvtilde}\\
&&\tilde{\bm w}_i(s)  \equiv \dot{\bm w}_i- \dot{\bm w}_{2n}.
\label{defwtilde}
\end{eqnarray}

For numerical purposes (\ref{lagtilde}) is discretized in time using
the standard Euler--Ito scheme of order one half~\cite{KP92}
\begin{equation} 
\tilde{\bm a}_i(s+\Delta s)-\tilde{\bm a}_i(s)=
\sqrt{\Delta s}\ ({\cal V}_i+\sqrt{2\kappa}\ {\cal W}_i),
\label{discrete} 
\end{equation}
where $\Delta s$ is the time step and the~${\cal V}_i$'s and
the~${\cal W}_i$'s ($i=1,\ldots,2n-1$) are $d$-dimensional Gaussian
random vectors chosen independently of each other and independently at
each time step and having the appropriate correlations, which are
calculated from (\ref{wn}) and (\ref{correlations}), namely
\begin{eqnarray}
&&  \left\langle {\cal V}_{i,\alpha}\,{\cal V}_{j,\beta} \right\rangle
=\nonumber\\
&&D_{\alpha\beta}(\tilde{\bm a}_i)+
D_{\alpha\beta}(\tilde{\bm a}_j) -D_{\alpha\beta}(\tilde{\bm a}_i
-\tilde{\bm a}_j) \ ,\label{dvcor} \\
&& \left\langle {\cal W}_{i,\alpha}\,{\cal
W}_{j,\beta} \right\rangle
= (1+\delta_{ij})\delta_{\alpha\beta},\label{dwcor} \end{eqnarray}
where $D_{\alpha\beta}({\bm r})$ is defined in (\ref{d-tensor}).

To actually generate these Gaussian random variables, we use the
symmetry and positive definite character of covariance matrices
like~(\ref{dvcor}) and~(\ref{dwcor}).  Indeed, any such matrix can be
factorized as a product (taking $\cal V$ as an example)~\cite{RR78}
\begin{equation} \langle {\cal V} \otimes {\cal V} \rangle = {\cal L}
{\cal L}^T \ , \label{choldec} \end{equation} where $\cal L$ is a
nonsingular lower triangular matrix and ${\cal L}^T$ is its transpose.
$\cal L$ can be computed explicitly using the Cholesky decomposition
method~\cite{RR78}, an efficient algorithm to compute the triangular
factors of positive definite matrices. It nevertheless  takes
$O\!\left([(2n-1)d]^3/3\right)$ flops to get~$\cal L$ from~$\langle
{\cal V} \otimes {\cal V} \rangle$ and this is the most time-consuming
operation at each time  step.  Once $\cal L$ is obtained, a suitable set of
variables~$\cal V$ can be obtained by applying the linear
transformation \begin{equation} {\cal V} = {\cal L}\;{\cal N}
\label{normal} \end{equation} to a set of {\em independent\/}
unit-variance Gaussian random variables~${\cal N}_{i,\alpha}$ coming
from a standard Gaussian random number generator, that is
with~$\langle {\cal N}_{i,\alpha}\,{\cal N}_{j,\beta} \rangle =
\delta_{ij}\delta_{\alpha\beta}$.  The resulting
variables~$({\cal V}_i)_\alpha$ then have the required covariances.

From the $\tilde{\bm a}_i(s)$'s we obtain the quantities $T_{ij}(L)$
for all the desired values of the integral scale $L$, typically, a
geometric progression up to the maximum value $L_{\rm max}$.  The
easiest way to evaluate~$S_{2n}(r;L)$ is to evolve simultaneously the
$n+1$~configurations corresponding to $p=0,\ldots,n$, stopping the
current realization when {\em all\/} inter-particle distances in {\em
all\/} configurations are larger than some appropriate large-scale
threshold~$L_{\rm th}$, to which we shall come back. The various
moments appearing in the structure functions are then calculated using
expression such as (\ref{c4}) in which the $(v,w)$-averaging is done
by the Monte-Carlo method, that is over a suitably large number of
realizations. We note that the expressions for the structure functions
of order higher than two involve heavy cancellations between the terms
corresponding to different configurations of particles.  For instance,
the three terms appearing in the expression (\ref{examples4}) for the
fourth-order function, all have dominant contributions scaling
as~$L^{2(2-\xi)}$ for large $L$ and a first subdominant correction
scaling as $L^{2-\xi}$. The true non-trivial scaling $\propto
L^{\zeta^{\rm anom}_4}$ emerges only after cancellation of the
dominant and first subdominant contributions.  For small $\xi$ the
dominant contributions are particularly large.  In the presence of
such cancellations, Monte-Carlo averaging is rather difficult since the
relative errors on individual terms decrease only as the inverse
square root of the number of realizations. In practice, the number of
realizations is increased until clean non-spurious scaling emerges. In
three dimensions, for ~$S_4(r;L)$, between one and several millions
realizations (depending on the value of~$\xi$) are required. In two
dimensions even more realizations are needed. For example, to achieve
comparable quality of scaling for $\xi=0.75$, in three dimensions
$4\times 10^6$ realizations are needed but $14\times 10^6$ are
needed in two dimensions.

Now, some comments on the choice of parameters. 

The threshold~$L_{\rm th}$ must be taken sufficiently large compared
to largest integral scale of interest~$L_{\rm max}$ to ensure that the
probability of returning within~$L_{\rm max}$ from~$L_{\rm th}$ is
negligible. But choosing an excessively large~$L_{\rm th}$ is too
demanding in computer resources.  In practice, the choice of~$L_{\rm
th}$ depends both on the space dimension and on how far one is from
the limit~$\xi= 0$.  In three dimensions, it is enough to take $L_{\rm
th} = 10\,L_{\rm max}$. In two dimensions there is a new difficulty
when $\xi$ is small. At $\xi=0$ the motion of Lagrangian particles and
also of separations of pairs of particles is exactly two-dimensional
Brownian motion. As it is well-known, in two dimensions, Brownian
motion is recurrent (see, e.g., Ref.~\cite{Fell}). Hence, with
probability one a pair of particles will eventually achieve
arbitrarily small separations. As a consequence, at $\xi=0$ in two
dimensions, the mean square value of $\theta$ is infinite. For very
small positive $\xi$ this mean square saturates but most of the
contribution comes from scales much larger than the integral scale.
This forces to choose extremely large values of ~$L_{\rm th}$ when
$\xi$ is small. In practice, for $0.6\le\xi\le0.9$ we take $L_{\rm th}
= 4 \times 10^3\;L_{\rm max}$ and beyond $\xi=0.9$ we take $L_{\rm
th}=10\,L_{\rm max}$. (The range $\xi<0.6$ has not yet been
explored.) In view of the accuracy of our results, we have verified
that the use of larger values for~$L_{\rm th}$ does not  affect in any
significant way the values of the scaling exponents.

The molecular diffusivity $\kappa$ is chosen in such a way that $r$ is
truly in the inertial range, namely, we demand (i) that the
dissipation scale $\eta \sim \kappa ^{1/\xi}$ should be
much smaller than the separation $r$ and (ii) that the time a pair
of particle spends with a separation comparable to $\eta$, which is
$\sim \eta ^2/\kappa \sim \eta ^{2-\xi}$  should be much smaller than 
the time needed for this separation to grow from $r$ to $L$, which
is $\sim L^{2-\xi} -r^{2-\xi}$. The latter condition becomes very
stringent
when $\xi$ is close to 2. 

Finally, the time step $\Delta s$ is chosen small compared to the
diffusion time $\eta ^2/\kappa$ at scale $\eta$ .

\section{Results}
\label{sec:res}

We now present results for structure functions up to fourth order.
The three-dimensional results have already been published in
Ref.~\cite{FMV98}.  The two-dimensional results are new. Some results
for structure functions of order six have been published in
Ref.~\cite{FMV-egs} and shall not be repeated here (more advanced 
simulations are in progress).

\begin{figure}
\begin{center}
\vspace{-0.6cm}
\mbox{\hspace{0.0cm}\psfig{file=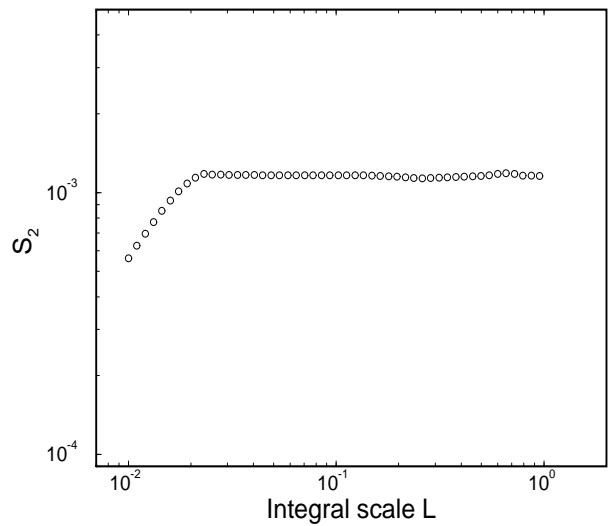,height=8cm,width=9cm}}
\end{center}
\vspace{-0.6cm}
\caption{3-D second-order structure function
$S_2$ {\it vs} $L$ for $\xi=0.6$. Separation $r=2.7\times 10^{-2}$,
diffusivity $\kappa=1.115\times 10^{-2}$, number of realizations
$4.5\times 10^6$.}
\label{fig3d06s2}
\end{figure}
A severe test for the Lagrangian method is provided by the
second-order structure function $S_2(r;L)$, whose expression is known
analytically \cite{K94}.  Its behavior being non-anomalous, a flat
scaling in $L$ should be observed. The $L$-dependence of $S_2$,
measured by the Lagrangian method, is shown in Fig.~\ref{fig3d06s2} for
$\xi=0.6$ and $d=3$ (all structure functions are plotted in log--log
coordinates).  The measured slope is $10^{-3}$ and the error on the
constant is $3\%$. (These figures are typical also for other values of
$\xi$ studied.)  We observe that, for separation $r$ much larger than
the integral scale $L$,  correlations between
$\theta({\bm r})$ and $\theta({\bm 0})$  are very small; hence, 
the  scaling for the second-order structure function is essentially
given by the $L$-dependence of  $\langle\theta^2\rangle$,
namely $L^{2-\xi}$; the transition to the constant-in-$L$ behavior
around $r=L$ is very sharp, on account of the step function chosen for
$F$.
\begin{figure}
\begin{center}
\mbox{\hspace{0.0cm}\psfig{file=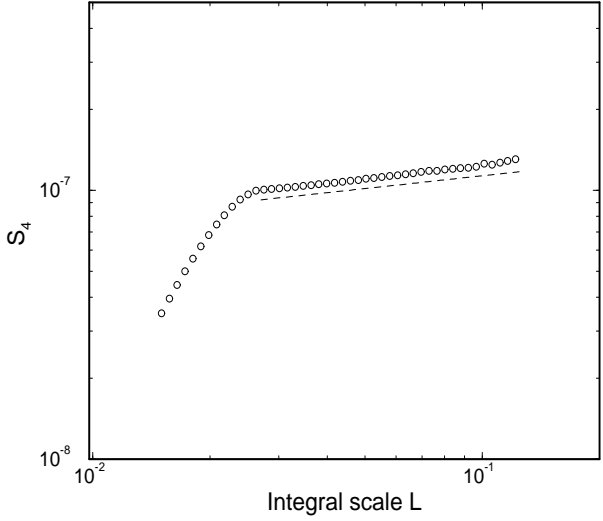,height=8cm,width=9cm}}
\end{center}
\vspace{-0.5cm}
\caption{3-D fourth-order structure function $S_4$ {\it vs} $L$ for
$\xi=0.2$. Separation $r=2.7\times 10^{-2}$, 
diffusivity $\kappa=0.247$, number of realizations  $15\times 10^6$.}
\label{fig3d02s4}
\end{figure}
\begin{figure}
\begin{center}
\mbox{\hspace{0.0cm}\psfig{file=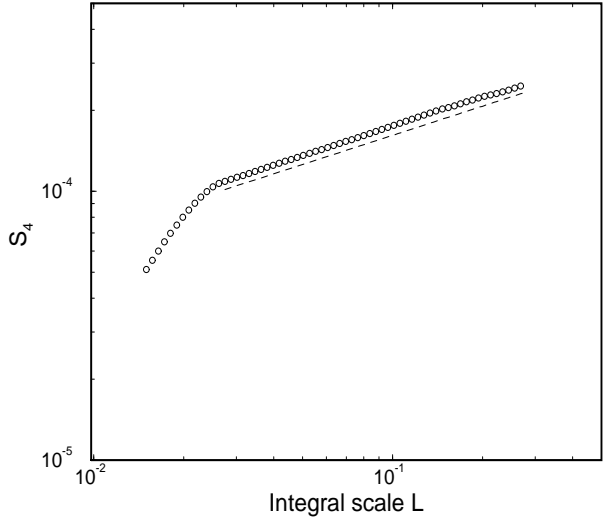,height=8cm,width=9cm}}
\end{center}
\vspace{-0.5cm}
\caption{Same as in Fig.~\protect\ref{fig3d02s4} for
$\xi=0.9$. Parameters\,:  $r=2.7\times 10^{-2}$, $\kappa=4.4\times
10^{-4}$,  number of realizations $8\times 10^6$.}
\label{fig3d09s4}
\end{figure}

\begin{figure}
\begin{center}
\mbox{\hspace{0.0cm}\psfig{file=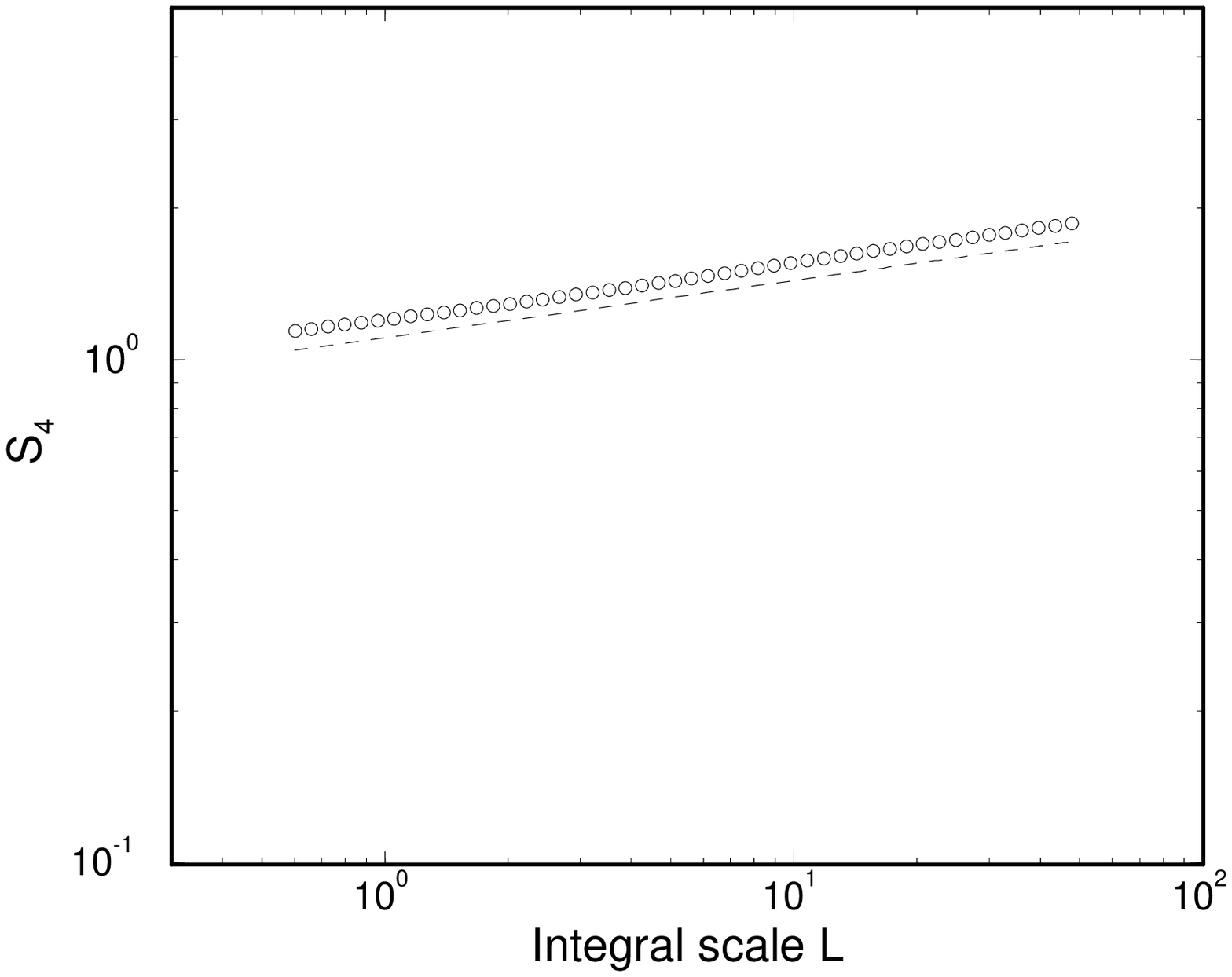,height=8cm,width=9cm}}
\end{center}
\vspace{-0.5cm}
\caption{Same as in Fig.~\protect\ref{fig3d02s4} for $\xi=1.75$.
Parameters\,: $r=2.7\times 10^{-2}$, $\kappa=10^{-9}$, number
of realizations $1.5\times 10^6$.}
\label{fig3d175s4}
\end{figure}

\begin{figure}
\begin{center}
\mbox{\hspace{0.0cm}\psfig{file=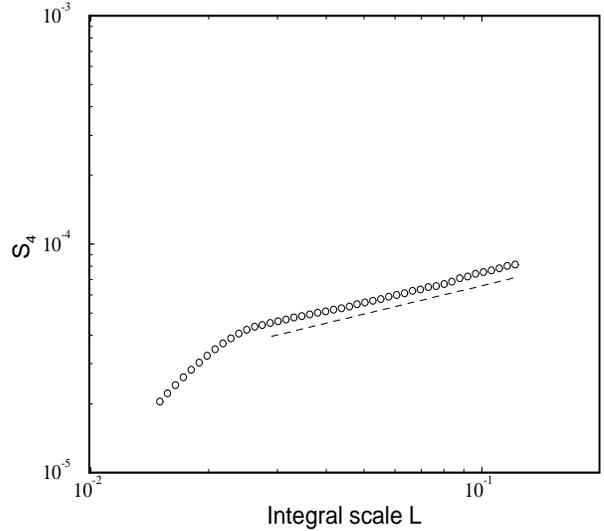,height=8cm,width=9cm}}
\end{center}
\vspace{-0.5cm}
\caption{2-D fourth-order structure function $S_4$ {\it vs} $L$ for
$\xi=0.6$. Parameters\,: 
$r= 2.7\times 10^{-2}$, $\kappa=1.1\times 10^{-2}$, number of
realizations  $ 5 \times 10^{6}$.}
\label{fig2d06s4}
\end{figure}

\begin{figure}
\begin{center}
\mbox{\hspace{0.0cm}\psfig{file=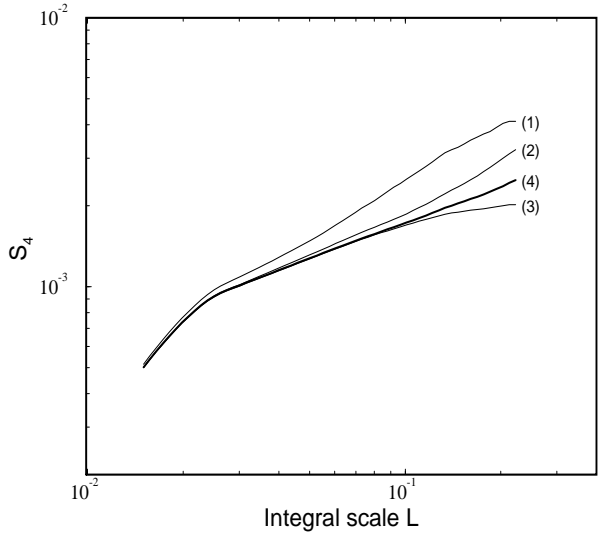,height=8cm,width=9cm}}
\end{center}
\vspace{-0.5cm}
\caption{Same as in Fig.~\protect\ref{fig2d06s4} for
$\xi=0.9$. Parameters are $r= 2.7\times 10^{-2}$, $\kappa=4.4\times
10^{-4}$. To illustrate convergence, various numbers of realizations
are shown\,: (1) $150\times 10^{3}$, (2) $1.5\times 10^{6}$, (3)
$3.4\times 10^{6}$, (4) from $4.8\times
10^{6}$ to  $7\times 10^{6}$ (several curves superposed).}
\label{fig2d09s4conv}
\end{figure}

\begin{figure}
\begin{center}
\mbox{\hspace{0.0cm}\psfig{file=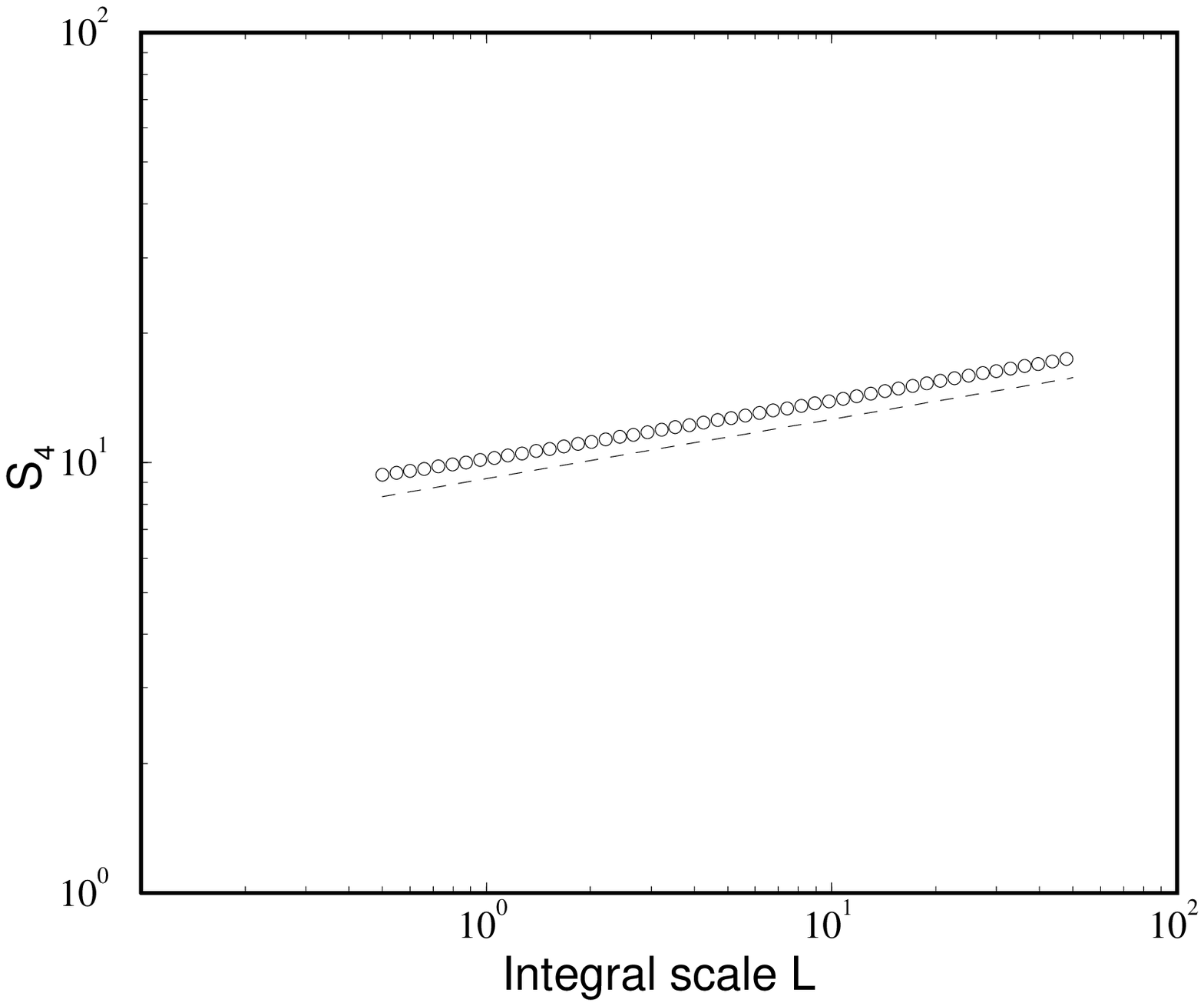,height=8cm,width=9cm}}
\end{center}
\vspace{-0.5cm}
\caption{Same as in Fig.~\protect\ref{fig2d06s4} for $\xi=1.75$. 
Parameters\,: $r=2.7\times 10^{-2}$,
$\kappa=10^{-9}$, number of realizations 
$2.4\times 10^6$.}
\label{fig2d175s4}
\end{figure}

Figs.~\ref{fig3d02s4}, \ref{fig3d09s4} and \ref{fig3d175s4} show the
$L$-dependence of the fourth-order structure function in three
dimensions for $\xi=0.2$, $0.9$ and $1.75$, respectively. In each case
the scaling region (which is basically $L>r$) is indicated by a dashed
straight line whose slope is the anomaly. Note that, to obtain a
similar high-quality scaling as shown on these figures, a much larger
number of realizations is needed for small $\xi$; this is required  to permit
cancellation of leading contributions to (\ref{examples4}), as
explained in Section~\ref{sec:num}.

The two-dimensional case, which is numerically more difficult for
reasons explained near the end of Section~\ref{sec:num}, is shown in
Figs.~\ref{fig2d06s4}, \ref{fig2d09s4conv} and \ref{fig2d175s4} for
$\xi=0.6$, $0.9$ and $1.75$, respectively. Fig.~\ref{fig2d09s4conv}
also shows the data obtained for various values of the number of
realizations. Note that if only $150\times 10^{3}$ realizations are
used, the anomaly (that is the slope obtained, e.g., by a least square
fit) is grossly overestimated.

\begin{figure}
\begin{center}
\mbox{\hspace{0.0cm}\psfig{file=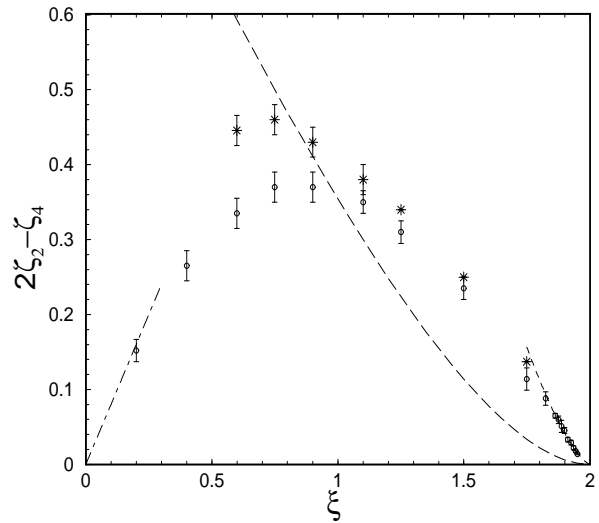,height=8cm,width=9cm}}
\end{center}
\vspace{-0.5cm}
\caption{The anomaly $2\zeta_2-\zeta_4$ for the fourth-order structure
function in  two dimensions (stars, upper graph) and three
dimensions (circles, lower graph). Error bars in 2-D shown only for $\xi \le
1.1$. The dashed line is the three-dimensional  linear ansatz prediction 
(\protect\ref{linearanom}).}
\label{fig2d3danom}
\end{figure}

Fig.~\ref{fig2d3danom} shows a plot of the anomaly $\zeta
^{\rm anom}_4$ {\it vs} $\xi$ in both two and three dimensions. The error
bars (shown in 2-D only for $\xi \le 1.1$ to avoid crowding) are obtained
by analyzing the fluctuations of local scaling exponents over octave
ratios of values for $L$, a method which tends to overestimate errors.

Let us now comment on the results.  In three dimensions, the error
bars near $\xi=2$ are exceedingly small and the data have a good
fit (shown as dashed line) of the form $\zeta ^{\rm anom}_4= a \gamma
+ b \gamma^{3/2}$ with $\gamma = 2-\xi$ (the parameters are $a=0.06$
and $b=1.13$). This is compatible with an expansion in powers of
$\sqrt{\gamma}$ \cite{SS96} in which a term $\propto\sqrt{\gamma}$ is
ruled out by the H\"older inequality $\zeta_4 \leq 2\zeta_2=2\gamma$.

Of particular significance is that, when $\xi$ is decreased from 2 to
0, the anomaly grows at first, achieves a maximum and finally
decreases. In three dimensions, where small-$\xi$ simulations are
easier than in two dimensions, we have good evidence that the anomaly
vanishes for $\xi\to 0$ as predicted by the perturbation theory of
Gaw\c{e}dzki and Kupiainen \cite{GK95}, whose leading order $\zeta
^{\rm anom}_4 = 4\xi/5$ is shown as a dot-dashed straight line on
Fig.~\ref{fig2d3danom}. Note that the next-order correction $\propto
\xi ^2$ is known \cite{AAV98} but the convergence properties of the
$\xi$-series are not clear. The ``linear ansatz'' prediction for the
anomaly given in Refs.~\cite{K94,KYC95}, that is
\begin{eqnarray}
&&\zeta ^{\rm anom}_4 ={3\zeta_2
+d\over2}-{1\over2}\sqrt{8d\zeta_2+(d-\zeta_2)^2},\label{linearanom}\\
&& \zeta_2 = 2-\xi,
\end{eqnarray}
is consistent with our results only near $\xi=1$, the point farthest
from the two limits $\xi=0$ and $\xi=2$, which both have strongly
nonlocal dynamics. This suggests a possible relation between
deviations from the linear ansatz and locality of the interactions
\cite{RHK98}. Whether this consistency for $\xi=1$ persists for
moments of order higher than four is an open problem.

The fact that the anomalies are stronger in two than in three
dimensions is consistent with their vanishing as $d\to \infty$
\cite{CFKL95}. The fact that the maximum anomaly occurs for a value
of $\xi$ smaller in two than in three dimensions can be tentatively
interpreted as follows. Near $\xi =0$ the dynamics is dominated by
the nearly ultraviolet-divergent eddy diffusion, whereas near $\xi=2$
it is dominated by the nearly infrared-divergent stretching. The
former increases with $d$, but not the latter. The maximum is achieved
when these two effects balance. 
 
\section{Concluding remarks}
\label{sec:conclusion}

Compared to Eulerian simulations of the partial differential equation
(\ref{passive}) our Lagrangian method has various advantages. When
calculating moments of order $2n$ we do not require the complete
velocity field at each time step but only $2n-1$ random vectors,
basically, the set of velocity differences between the locations of
Lagrangian tracer particles which are advected by the flow and
subject to independent Brownian diffusion. As a consequence the
complexity of the computation (measured in number of floating point
operations) grows polynomially rather than exponentially with the
dimension $d$. Furthermore, working
with tracers naturally allows to measure the scaling of the structure
functions $S_{2n}(r;L)$ {\it vs} the integral scale $L$ of the
forcing. Physically, this means that the injection rate of passive
scalar variance (which equals its dissipation rate) and the separation
$r$ are kept fixed while the integral scale $L$ is varied.  Anomalies,
that is discrepancies from the scaling exponents which would be
predicted by na\"{\i}ve dimensional analysis, are measured here
directly through the scaling dependence on $L$ of the structure
functions.

Actually, direct Eulerian simulations and the Lagrangian method are
complementary. Very high-resolution simulations of the sort found in
Ref.~\cite{CK98} are really not practical in more than two
dimensions but do give access to the entire Eulerian passive scalar
field. Hence, they can  and have been used to address questions
about the geometry of the scalar field and about probability
distributions.

Although we have presented here the numerical implementation of the
Lagrangian method only  for the Kraichnan model, it is clear that 
the general strategy presented in Sections~\ref{sec:lag} and
\ref{sec:kappazero} is applicable to a wide class of random flows,  for
example, with a finite correlation time.

An important aspect of the results obtained for the Kraichnan model is
that they agree with perturbation theory \cite{GK95} for $\xi\to
0$. As is now clear from theory and simulations, anomalies in the
Kraichnan model and other passive scalar problems arise from zero
modes in the operators governing the Eulerian dynamics of $n$-point
correlation functions \cite{GK95,CFKL95,SS95}. It is likely that some
form of zero modes is also responsible for anomalous scaling in
nonlinear turbulence problems.  An instance are the ``fluxless
solutions'' to Markovian closures based on the Navier--Stokes
equations in dimension $d$ close to two, which have a power-law
spectrum with an exponent depending continuously on $d$
\cite{FLSF}. Attempts to capture intermittency effects in
three-dimensional turbulence are being made along similar lines (see,
e.g., Ref.~\cite{BLPP98}).  The Kraichnan model for passive scalar
intermittency puts us on a trail to understanding anomalous scaling in
turbulence.\\

\par\noindent {\bf ACKNOWLEDGMENTS}

We are most grateful to Robert~Kraichnan for innumerable interactions
on the subject of passive scalar intermittency over many years. We
acknowledge useful discussions with M.~Chertkov, G.~Falkovich, O.~Gat,
K.~Gaw\c{e}dzki, F.~Massaioli, S.A.~Orszag, I.~Procaccia, A.~Wirth,
V.~Yakhot and R.~Zeitak.  Simulations were performed in the framework
of the SIVAM project of the Observatoire de la C\^ote d'Azur.  Part of
them were performed using the computing facilities of CASPUR at Rome
University, which is gratefully acknowledged. Partial support from the
Centre National de la Recherche Scientifique through a ``Henri
Poincar\'e'' fellowship (AM) and from the Groupe de Recherche
``M\'ecanique des Fluides G\'eophysiques et Astrophysiques'' (MV) is
also acknowledged.

\newpage
\noindent FIGURE CAPTIONS

\vspace{1mm}
\noindent Fig. 1. 3-D second-order structure function
$S_2$ {\it vs} $L$ for $\xi=0.6$. Separation $r=2.7\times 10^{-2}$,
diffusivity $\kappa=1.115\times 10^{-2}$, number of realizations
$4.5\times 10^6$.

\vspace{1mm}
\noindent Fig. 2. 3-D fourth-order structure function $S_4$ {\it vs} $L$ for
$\xi=0.2$. Separation $r=2.7\times 10^{-2}$, 
diffusivity $\kappa=0.247$, number of realizations  $15\times 10^6$.

\vspace{1mm}
\noindent Fig. 3 Same as in Fig.~2 for
$\xi=0.9$. Parameters\,:  $r=2.7\times 10^{-2}$, $\kappa=4.4\times
10^{-4}$,  number of realizations $8\times 10^6$.

\vspace{1mm}
\noindent Fig. 4. Same as in Fig.~2 for $\xi=1.75$.
Parameters\,: $r=2.7\times 10^{-2}$, $\kappa=10^{-9}$, number
of realizations $1.5\times 10^6$.

\vspace{1mm}
\noindent Fig. 5. 2-D fourth-order structure function $S_4$ {\it vs} $L$ for
$\xi=0.6$. Parameters\,: 
$r= 2.7\times 10^{-2}$, $\kappa=1.1\times 10^{-2}$, number of
realizations  $ 5 \times 10^{6}$.

\vspace{1mm}
\noindent Fig. 6. Same as in Fig.~5 for
$\xi=0.9$. Parameters are $r= 2.7\times 10^{-2}$, $\kappa=4.4\times
10^{-4}$. To illustrate convergence, various numbers of realizations
are shown\,: (1) $150\times 10^{3}$, (2) $1.5\times 10^{6}$, (3)
$3.4\times 10^{6}$, (4) from $4.8\times
10^{6}$ to  $7\times 10^{6}$ (several curves superposed).

\vspace{1mm}
\noindent Fig. 7. Same as in Fig.~5 for $\xi=1.75$. 
Parameters\,: $r=2.7\times 10^{-2}$,
$\kappa=10^{-9}$, number of realizations 
$2.4\times 10^6$.

\vspace{1mm}
\noindent Fig. 8. The anomaly $2\zeta_2-\zeta_4$ for the fourth-order structure
function in  two dimensions (stars, upper graph) and three
dimensions (circles, lower graph). Error bars in 2-D shown only for $\xi \le
1.1$. The dashed line is the three-dimensional  linear ansatz prediction 
(49).
\end{document}